\documentclass[twocolumn, resetfootnote, tighten]{aastex701}
\usepackage[
    separate-uncertainty=true,
    multi-part-units=single,
    range-phrase=--,
    range-units=single
]{siunitx}
\DeclareSIUnit{\au}{AU}
\DeclareSIUnit{\dex}{dex}
\DeclareSIUnit{\hjulianday}{HJD}
\DeclareSIUnit{\Jansky}{Jy}
\DeclareSIUnit{\julianday}{JD}
\DeclareSIUnit{\bjd}{BJD}
\DeclareSIUnit{\mas}{mas}
\DeclareSIUnit{\mag}{mag}
\DeclareSIUnit{\parsec}{pc}
\DeclareSIUnit{\permille}{\text{\textperthousand}}
\DeclareSIUnit{\rad}{rad}
\DeclareSIUnit{\solarmass}{M_\odot}
\DeclareSIUnit{\solarradius}{R_\odot}
\DeclareSIUnit{\solarluminosity}{L_\odot}
\DeclareSIUnit{\earthmass}{M_\oplus}
\DeclareSIUnit{\year}{yr}
\defcitealias{gaiaDR3}{GDR3}
\defcitealias{Baycroft2023}{B23}

\begin{document}

\title{Stellar activity demotes GJ~9404~b to a planetary candidate}

\author[orcid=0000-0002-6059-1178]{Atanas K. Stefanov}
\affiliation{Observatoire Astronomique de l’Universit\'e de Gen\`eve, 1290 Versoix, Switzerland}
\affiliation{Instituto de Astrof\'isica de Canarias, 38205 La Laguna, Spain}
\affiliation{Departamento de Astrof\'isica, Universidad de La Laguna, 38206 La Laguna, Spain}
\email[show]{atanas.stefanov@unige.ch}

\author[orcid=0000-0002-3814-5323]{A. Suárez Mascareño}
\affiliation{Instituto de Astrof\'isica de Canarias, 38205 La Laguna, Spain}
\affiliation{Departamento de Astrof\'isica, Universidad de La Laguna, 38206 La Laguna, Spain}
\email{alejandro.suarez.mascareno@iac.es}

\author[orcid=0000-0002-0264-7356]{Jonay I. González Hernández}
\affiliation{Instituto de Astrof\'isica de Canarias, 38205 La Laguna, Spain}
\affiliation{Departamento de Astrof\'isica, Universidad de La Laguna, 38206 La Laguna, Spain}
\email{jonay.gonzalez@iac.es}

\author[orcid=0009-0004-7844-9729]{P. Charpentier}
\affiliation{Instituto de Astrof\'isica de Canarias, 38205 La Laguna, Spain}
\affiliation{Departamento de Astrof\'isica, Universidad de La Laguna, 38206 La Laguna, Spain}
\email{paul.charpentier@iac.es}

\author[orcid=0009-0001-2825-6185]{N. Nari}
\affiliation{Light Bridges S. L., 35004 Las Palmas de Gran Canaria, Spain}
\affiliation{Instituto de Astrof\'isica de Canarias, 38205 La Laguna, Spain}
\affiliation{Departamento de Astrof\'isica, Universidad de La Laguna, 38206 La Laguna, Spain}
\email{nicola.nari@lightbridges.es}

\author[orcid=0000-0003-3767-7085]{R. Rebolo}
\affiliation{Instituto de Astrof\'isica de Canarias, 38205 La Laguna, Spain}
\affiliation{Departamento de Astrof\'isica, Universidad de La Laguna, 38206 La Laguna, Spain}
\affiliation{Consejo Superior de Investigaciones Cient\'ificas (CSIC), 28006 Madrid, Spain}
\email{rrl@iac.es}

\author[orcid=0000-0002-3300-3449]{T. A. Baycroft}
\affiliation{Tsung-Dao Lee Institute, Shanghai Jiao Tong University, 201210 Shanghai, China}
\email{txb187@student.bham.ac.uk}

\begin{abstract}
GJ~9404 is a nearby M dwarf \mbox{($d=23.8\,\unit\parsec$)}, around which a \mbox{$13.4586^{+0.0044}_{-0.0067}\,\unit\day$} planet has been claimed in a recent study using 54 publicly available HARPS-N radial velocities (RVs). We re-analyse the spectroscopy associated with this study, and carry out a joint modelling of: (i)~stellar rotation in RV, FWHM, and Mount Wilson S-index through Gaussian processes; (ii)~long-term trends in all physical quantities; (iii)~planetary signals in RV. Current evidence supports the existence of planets in the period range \SIrange{2}{100}{\day} by at most \mbox{$\Delta\ln Z=-0.4$} for any considered stellar-activity model, which is insufficient for a widely accepted planet detection. Additional spectroscopy is necessary to accurately quantify the stellar activity of GJ~9404 and determine whether any RV signals warrant further attention.
\end{abstract}



\section{Introduction}

Although M dwarfs remain the theoretical favourites for exoplanetary radial-velocity (RV) detection by virtue of their low masses, in practice, they often attenuate or imitate planetary RV signals through short- and long-term stellar activity (e.g. \citealp{Borgniet2015} and \citealp{Dumusque2011a}, respectively). We investigate the RV behaviour of one such nearby M dwarf, GJ~9404
\mbox{($\alpha=12\textsuperscript{h}19\textsuperscript{m}24.1\textsuperscript{s}$},
\mbox{$\delta=28\degr22'56.5''$},
\mbox{$d=\SI{23.8}{\parsec}$}; \citealp{gaiaDR3}).
GJ~9404 has been studied by the HADES RV programme \citep{Affer2016} that was conducted at the HARPS-N spectrograph \citep{harpsn}. \citet{Pinamonti2022} proposed a planetary candidate around GJ~9404 with an orbital period of
\mbox{$P=13.46^{+0.01}_{-0.51}\,\unit\day$}
and a minimum mass of
\mbox{$m\sin i=10.3\pm 1.8\,\unit\earthmass$}. Later, using the same data, \citet{Baycroft2023}, hereafter \citetalias{Baycroft2023}, declared the candidate to be a confirmed planet: GJ~9404~b, with an orbital period of
\mbox{$P=13.4586^{+0.0044}_{-0.0067}\,\unit\day$},
a minimum mass of
\mbox{$m\sin i=11.9\pm 1.9\,\unit\earthmass$},
and an eccentricity of
\mbox{$e=0.49^{+0.11}_{-0.13}$}.
However, their analysis considered RVs only, and did not account for stellar activity, which is known to produce false positives (e.g. \citealp{Queloz2001, Robertson2014,Rajpaul2015a,Bortle2021}).
In this work, we aim to determine whether current evidence supports the significance of GJ~9404~b, having accounted for stellar activity.

\section{Data}
We reduced the historic HARPS-N spectra through the YABI portal hosted by the \textit{Centro Italiano Archivi Astronomici}. YABI \citep{yabi1,yabi2} implements the classical HARPS-N pipeline and computes cross-correlation functions (CCFs) for a given template mask. We used an M2 mask to obtain CCFs. Then, we fed all raw spectra to the line-by-line method of \citet{Artigau2022},
with an effective-temperature estimate informed by \citet{Maldonado2021}: $T_\text{eff}=\SI{3891(70)}{\kelvin}$. This yielded 54 RVs with a root mean square (rms) of \SI{4.08}{\metre\per\second}, a median uncertainty of \SI{1.00}{\metre\per\second}, and a baseline of \SI{1432}{\day}. \citet{hires_survey} contains 6 HIRES measurements of GJ~9404. We did not consider them on account of being too few and distant from the HARPS-N time series.

Active regions perturb the flux- and velocity fields of the stellar disc, thereby altering the shape of observed lines. Consequently, the CCF also changes form. CCF FWHMa are automatically provided by the \mbox{HARPS-N} data reduction software; we took them and colour-corrected them order by order, following \citet{SuarezMascareno2023}. The final FWHM time series are described by an rms of \SI{8.56}{\metre\per\second} and a median uncertainty of \SI{3.91}{\metre\per\second}.

The emission intensity of the cores of Ca II H\&K lines is a reliable proxy for the strength of the stellar magnetic field, and thereby of the stellar-rotation period $P_\text{rot}$ for late-type stars \citep{Noyes1984,Lovis2005}. We used the Mount Wilson \mbox{S-index}, $S_\text{MW}$, as defined in \citet{Vaughan1978}. These time series are described by an rms of \num{0.149} and a median uncertainty of \num{0.071}.

\section{Methods}\label{sec:methods}
We employ the modelling framework described in \citet{Stefanov2025a}, which includes, but is not limited to:
(i)~long-term functions (LTFs) for physical or instrumental trends,
(ii)~source-dependent jitters for each quantity,
(iii)~elliptical- and circular-orbit fitting,
(iv)~stellar-activity modelling using Gaussian processes (GPs). We utilised multi-dimensional GPs in the $FF'$ framework \citep{Aigrain2012, Rajpaul2015}. For the stellar-activity component of measurements $y$ at time $t$, we solved the system of equations
\begin{equation}
    \left|\begin{array}{l}
    y_\text{RV}=A_\text{RV}\,G(t)+B_\text{RV}\,(\mathrm{d}G/\mathrm{d}t)_{t}\\[0.2ex]
    y_\text{FWHM}=A_\text{FWHM}\,G(t)\\[0.2ex]
    y_{S_\text{MW}}=A_{S_\text{MW}}\,G(t)
    \end{array}\right.
    \label{eq:multidimensional_system}
\end{equation}
where $A_j$ and $B_j$ scale stellar activity in different quantities, and $G$ is its latent-kernel estimate. We used the \textsc{s+leaf} MultiSeriesKernel procedure to realise this framework \citep{spleaf1, spleaf2}.
We utilised \mbox{ReactiveNestedSampler}, a nested-sampling integrator provided by \textsc{UltraNest} \citep{ultranest}, to numerically evaluate the Bayesian evidence $Z$. In every inference instance of $N_\text{param}$ model parameters, we required $40N_\text{param}$ live points and a region-respecting slice sampler that accepted $4N_\text{param}$ steps until the sample was considered independent.

\section{Analysis and results}
We inspected the generalised Lomb-Scargle periodograms (GLSPs; \citealp{Lomb1976, Scargle1982, Zechmeister2009, VanderPlas2015}) of RV, FWHM and S-index in the period range \SIrange{2}{3000}{\day}. The RV time series are devoid of strong powers at long periods. There is a clustering of signatures between \SIrange{10}{20}{\day}, most notably at \SI{13.5}{\day}, \SI{15.9}{\day}, and \SI{17.2}{\day}. The FWHM and S-index GLSPs are somewhat similar, with significant long-term peaks near \SI{260}{\day} and near \SI{1000}{\day}. In terms of short-term activity, we observe strong peaks at \SI{37.1}{\day} and \SI{39.9}{\day} in FWHM; as well as at \SI{20.0}{\day}, \SI{33.2}{\day}, and \SI{81.5}{\day} in S-index. Some of these peaks are near the measured rotational period \mbox{$P_\text{rot}=23.2\pm 0.1\,\unit\day$} by \citet{SuarezMascareno2018}.

We proceeded to choose an appropriate LTF before including a GP component. The evidence favoured a model with zero-order trends in each dimension and a sinusoid of common period. We used the following priors for said sinusoidal:
\mbox{$\mathcal{U}_{\log}(200,2000)\,\unit\day$}
for the common period,
\mbox{$\mathcal{U}(0,20)\,\unit{\metre\per\second}$}
for the cycle amplitudes in RV and FWHM, and
\mbox{$\mathcal{U}(0,1)$}
for the S-index amplitude. We obtained a dominant posterior with contributions in all dimensions at the common period $262\pm2\,\unit\day$ (96\% of weight), as well as with a weak mode at $1043^{+22}_{-15}\,\unit\day$ (3\% of weight). In the residual periodogram, all signatures beyond \SI{100}{\day} effectively disappeared. We favoured the first solution by virtue of its weight and its high repeatability relative to the \SI{1432}{\day} baseline.

Then, we included a multi-dimensional GP within an $FF'$ framework, as described earlier. We invoked three approximations of the quasi-periodic kernel:
\begin{equation}
    k(\Delta t) = A^2\exp\left[
    -\frac{\Delta t^2}{2\tau^2}
    -\frac{\sin^2\left(\pi\Delta t/P\right)}{2\eta^2}
    \right],
    \label{eq:sep_kernel}
\end{equation}
in particular, the Matérn 3/2 exponential periodic (MEP) kernel, and the exponential-sine periodic kernel with two and three harmonics (ESP2, ESP3), all as defined in \textsc{s+leaf}. We used the following priors: a common-cycle period of
\mbox{$\mathcal{U}(200,300)\,\unit\day$},
a timescale of coherence of
\mbox{$\mathcal{U}_{\log}(20,10^4)\,\unit\day$},
a stellar-rotation period of
\mbox{$\mathcal{N}(23,1)\,\unit\day$},
and a harmonic complexity of
\mbox{$\mathcal{U}_{\log}(0.1,3)$}. Additionally, we searched agnostically for a circular- or an elliptic-orbit planet, with the following priors:
an orbital period of
\mbox{$\mathcal{U}_{\log}(2,100)\,\unit\day$},
a semi-amplitude of
\mbox{$\mathcal{U}(0,10)\,\unit{\metre\per\second}$},
and in the case of elliptic orbits,
an eccentricity of
\mbox{$\mathcal{U}(0,1)$}.
We compared these three quasi-periodic kernels against an LTF-only model that did not include a GP kernel. This resulted in a \numproduct{4x3} model grid that probed different stellar-activity covariance assumptions and planetary configurations.

We report the following $\ln Z$ values of LTF-only models, as well as those with MEP, ESP2, and ESP3 kernels:
\mbox{$-313.7$,} \mbox{$-281.3$,} \mbox{$-282.8$,} \mbox{$-282.9$} for planet-free models;
\mbox{$-305.3$,} \mbox{$-282.3$,} \mbox{$-283.2$,} \mbox{$-283.6$} for circular-orbit models; and
\mbox{$-305.6$,} \mbox{$-281.7$,} \mbox{$-283.9$,} \mbox{$-283.7$} for elliptic-orbit models. The evidence overwhelmingly favours models that account for correlated noise with a GP kernel \mbox{($\Delta\ln Z>24$)}, so we restrict our discussion to those. All planet-bearing GP models are disfavoured: $\Delta\ln Z \leq -0.4$, regardless of the kernel. This is well below the conventional threshold for statistical significance (\mbox{$\Delta\ln Z \geq 5$}; \citealp{Jeffreys1946}). 

\begin{figure*}[t]
\plotone{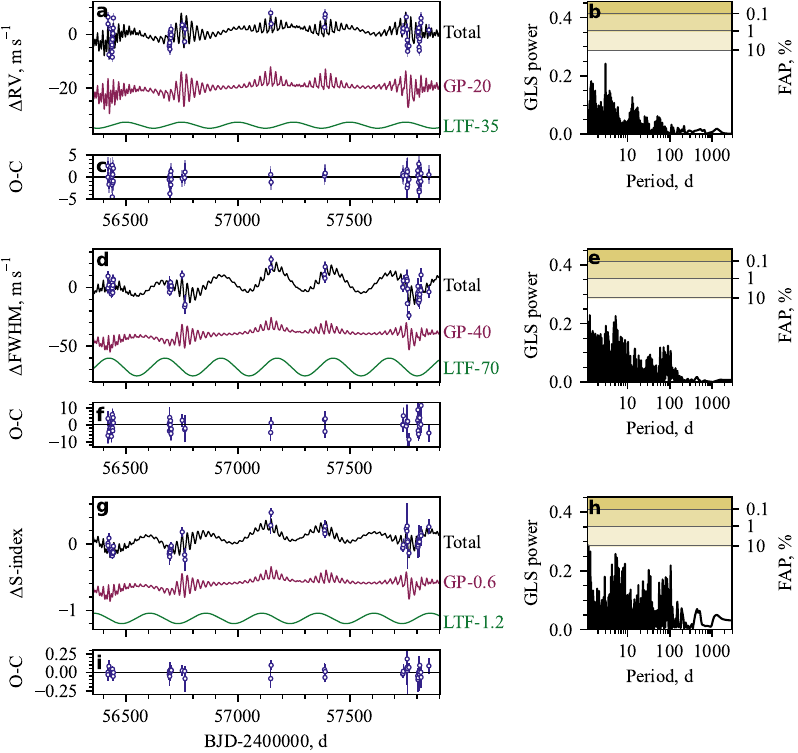}
\caption{
    Raw time series (blue points) against our best model (MEP, planet-free; solid lines) over the baseline, in (a)~RV, (d)~FWHM, (g)~S-index. Data points come with two error bars: one assigned by the instrument (blue), and another that includes model jitter (light blue). (c,f,i)~Associated residual time series. (b,e,h)~Associated GLSPs of residual time series.
    \vspace{1em}
}
\label{fig:bestmodel_fit}
\end{figure*}

Figure~\ref{fig:bestmodel_fit} displays the model prediction of the model elected by the evidence (planet-free MEP; hereafter `best model'), and shows the residual GLSPs, having accounted for model jitter.
Our best model recovered our informative prior on the rotation period
\mbox{$P_\text{rot}=23.6^{+1.1}_{-0.9}\,\unit\day$}. That is to say, the spectroscopic data neither corroborate the measurement of \citet{SuarezMascareno2018}, nor refute it. We measure a timescale of coherence
\mbox{$\tau =66^{+103}_{-33}\,\unit\day$}, the posterior shape of which becomes zero-converging in non-MEP models. We report a harmonic complexity of
\mbox{$\eta=0.77^{+0.27}_{-0.21}$}
and an RV jitter of
\mbox{$J_\text{rv}=1.65^{+0.37}_{-0.32}\,\unit{\metre\per\second}$}, both being compatible with other models in the grid.

All models that include a planetary signal catch an RV signature near \SI{13}{\day}, though with poorly-determined phases. The ESP2 eccentric-planet model assigned a zero-converging semi-amplitude, while the rest measured semi-amplitudes near
\mbox{$3\pm 1\,\unit{\metre\per\second}$}, smaller than what \citetalias{Baycroft2023} measured
\mbox{($5.13^{+1.04}_{-0.85}\,\unit{\metre\per\second}$)}. All elliptic-orbit models had poorly-determined posteriors in $e$ except the LTF-only model ($0.23^{+0.15}_{-0.14}$).

\section{Conclusions}
While the recovery of a \SI{13}{\day} RV signal from all uninformed searches is encouraging, current evidence does not support the inclusion of any planet around GJ~9404 as soon as a stellar-activity model is involved. We remark that just a few dozen measurements demonstrate the ability to reach an acceptable multi-dimensional $FF'$ stellar-activity model, as long as one uses two or more activity indicators. Nevertheless, additional spectroscopic data are needed to accurately determine the stellar-activity parameters of GJ~9404, and in particular, to validate its measured rotational period.

\newpage
\bibliography{zotero_bibtex}{}
\bibliographystyle{aasjournalv7}

\end{document}